\documentclass[a4paper,keeplastbox]{jacow}
\usepackage{pdfpages,multirow,ragged2e}
\makeatletter\ifboolexpr{bool{xetex}}
{\renewcommand{\Gin@extensions}{.pdf,%
    .png,.jpg,.bmp,.pict,.tif,.psd,.mac,.sga,.tga,.gif,%
    .eps,.ps,%
  }}{}
\makeatother
\ifboolexpr{bool{xetex} or bool{luatex}} 
 {}                             
 {\usepackage[utf8]{inputenc}}           

\usepackage[USenglish]{babel}
\usepackage{lineno}
\usepackage{fancyvrb} 

\begin{document}
\title{Design of a new SPS injection system via numerical optimisation}
\author{E. Waagaard\thanks{also at Uppsala University, Sweden}, M. J. Barnes, W. Bartmann, J. Borburgh, L. S. Ducimeti\`ere, T. Kramer, T. Stadlbauer,\\ P. Trubacova, F. M. Velotti, CERN, Geneva, Switzerland}
\maketitle
\begin{abstract}
The Super Proton Synchrotron (SPS) injection system plays a fundamental role to preserve the quality of injected high-brightness beams for the Large Hadron Collider (LHC) physics program and to maintain the maximum storable intensity. The present system is the result of years of upgrades and patches of a system not conceived for such intensities and beam qualities. In this study, we propose the design of a completely new injection system for the SPS using multi-level numerical optimisation, including realistic hardware assumptions. We also present how this hierarchical optimisation framework can be adapted to other situations for optimal accelerator system design.
\end{abstract}
%
\section{Introduction} 
\label{sec:Introduction}
\begin{figure}[htb!]
\centering
\includegraphics[width=0.99\columnwidth]{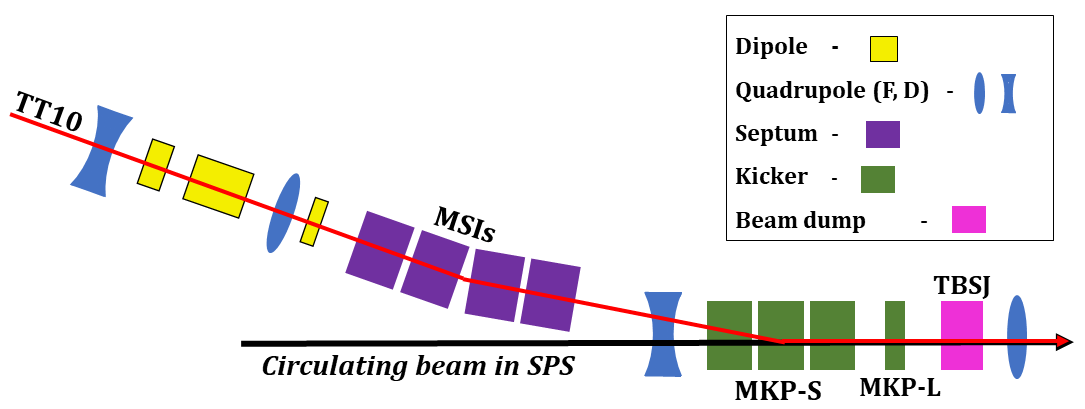}
\caption{\label{fig:sps_tt10_injection} Illustration of present TT10/SPS injection.}
\end{figure}

The SPS is the last injector of the LHC, hence the second-largest accelerator of the CERN accelerator complex~\cite{sps}. One of the main duties of the SPS is to provide high-brightness beams of high quality to the LHC physics program, as well as high-intensity beams to the fixed-target experiments of the North experimental Area (NA). The SPS injection system plays a fundamental role to the SPS beam quality, not only to preserve the quality of the injected beam, but also to maintain the maximum storable intensity in the ring. In particular, the upcoming High-Luminosity LHC~\cite{highlumi} requirements where the present injection system will act as a bottleneck for higher intensities further stresses the need for a new robust injection solution. Presently, the beam injected into the SPS is delivered from the Proton Synchrotron (PS) via the TT10 transfer line. The injection occurs in the horizontal plane in a single turn. The schematic view of the main elements is illustrated in Fig.~\ref{fig:sps_tt10_injection}. 

\begin{figure}[htb!]
\centering
\includegraphics[width=.8\columnwidth]{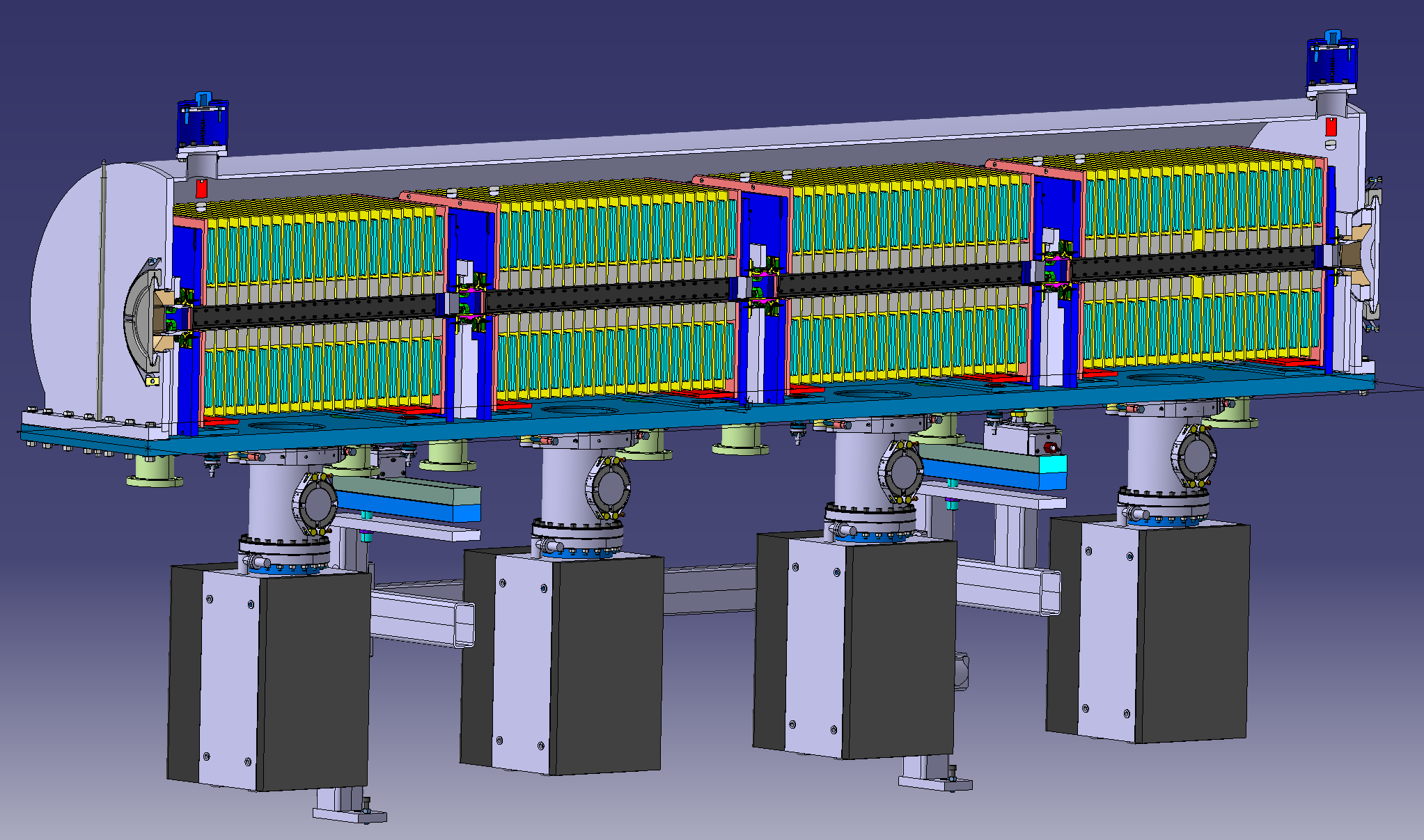}
\caption{\label{fig:mkpl_tank_xsection} Cross-section of MKP-L kicker tank.}
\end{figure}

There are four septa (denoted MSI) - insertion magnets that define the borderline between the ring and the transfer line (TL) - in the present injection system. Four kicker magnets - fast-pulsed dipole magnets - give the final deflection to the injected beam. There are two main types of kicker magnets: MKP-S and MKP-L. A cross-section of the MKP-L kicker vacuum tank is shown in Fig.~\ref{fig:mkpl_tank_xsection}. Due to difference in aperture, the MKP-L is more susceptible to beam-induced heating via beam coupling impedance. Thus, the new injection kickers should be designed also accounting for these constraints. In case of injection failure, a dedicated beam dump denoted TBSJ is used. 

\section{Injection simulations} 
\label{sec:injection_simulations}

For the studies and simulations, a model of the SPS injection system was developed using MAD-X~\cite{madx}, presented in detail in~\cite{elias_report}. To simplify the integration with other frameworks for optimisation, we used the python wrapper \Verb|cpymad|~\cite{cpymad}. The three main optics used during normal SPS operations are \textbf{Q20}, \textbf{Q26} and \textbf{SFTPRO}~\cite{sps_optics}. Q20 is used for beam delivery to LHC, whereas Q26 is the previous optics version for LHC but is now used together with Q20 for high-intensity bunched beams of LHC type. On the other hand, SFTPRO is used for slow extraction for the fixed-target experiments of NA. 


\section{Optimisation parameters}

\begin{figure}[htb!]
\centering
\includegraphics[width=\columnwidth]{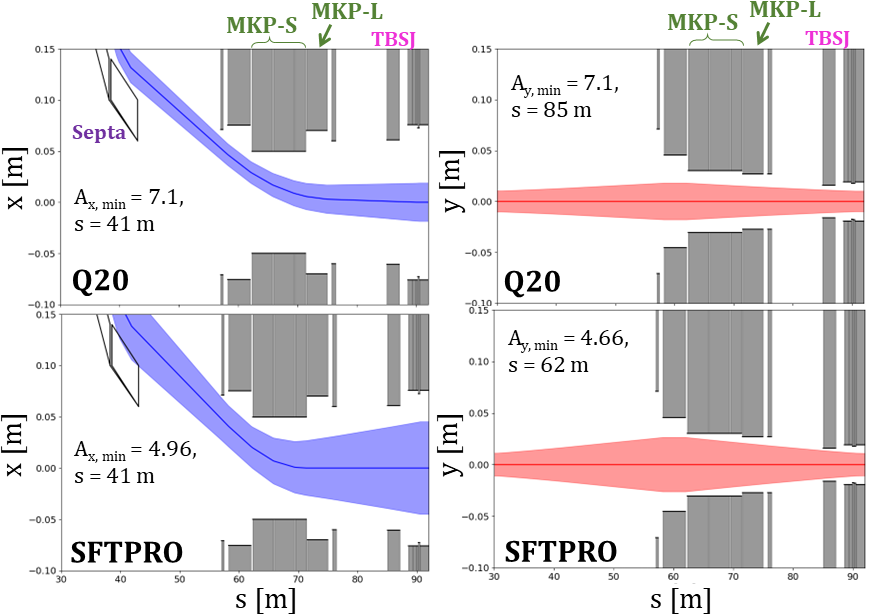}
\caption{\label{fig:todays_injection_envelope} MAD-X simulations of the current horizontal (left) and vertical (right) envelope injection for Q20 and SFTPRO optics, with the minimum acceptance $A_{\textrm{min}}$ for each case and at which longitudinal position $s$ it occurs.} 
\end{figure}

It is crucial to minimise errors at injection to avoid injection oscillations in the SPS, which ultimately can lead to filamentation and emittance blow-up, reducing luminosity and beam quality delivered to the experiments. Thus, we strive to minimise horizontal displacement at injection with zero-valued position $x$ and transverse momenta $x'$ for a stable beam. A safe beam dump is also required - when the MKPs are powered off, the beam has to hit the TBSJ. We also need to avoid any beam losses, which occur if the beam touches the mechanical aperture. For this purpose, the minimum acceptance $A_{x\textrm{, min}}$ (which is the number of standard deviations of the beam size that fits into the available space between the orbit and the mechanical aperture) is used as a figure of merit:

\begin{equation}
A_{x\textrm{, min}} = \textrm{min} \bigg( \frac{\textrm{aper}_x(s) - |x(s)|}{\sigma_x(s)} \bigg),     
\label{eq:min_acc}
\end{equation}

where $\textrm{aper}_x(s)$ is the horizontal mechanical aperture of the machine, with $y$ replacing $x$ for the vertical case. For the horizontal plane, the beam size ${\sigma_x(s)}$ is defined as 

\begin{equation}
\sigma_x(s) = \sqrt{\beta_x(s) \varepsilon_x + (D_x(s) \Delta p / p)^2}, 
\end{equation}

where $D_x$ is the dispersion function, $\varepsilon_x$ is the geometric emittance, $\beta_x$ is the beta function, and $ \Delta p / p$ is the momentum spread. MAD-X simulations of the horizontal and vertical beam envelopes for the present SPS injection system with corresponding $A_{x\textrm{, min}}$ values are shown in Fig.~\ref{fig:todays_injection_envelope}.

\begin{figure}[htb!]
\centering
\includegraphics[width=.75\columnwidth]{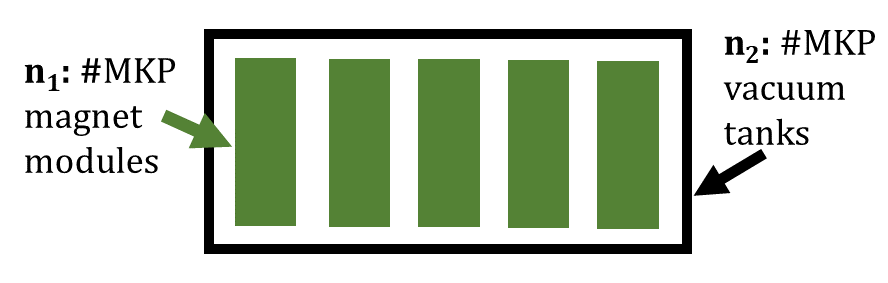}
\caption{\label{fig:mkp_modular_sketch} Illustration of the MKP magnet tank.}
\end{figure}

The strict requirements on a new injection system listed above - a stable preserved beam that can be safely dumped, applicable to all the optics - are primarily fulfilled by varying magnet strengths and positions of the septa and the kickers. At the lower level, these parameters constitute the degrees of freedom (DOF), or \textit{actors}. Once these low-level requirements are fulfilled, we also strive to deploy minimal magnet resources: maximum magnetic field $B$ required and integrated field $\int B dl$, for facilitated magnet design. For this reason, we can also at the higher level vary the number $n_1$ of MKP magnet modules per tank and the number $n_2$ of MKP vacuum tanks, illustrated in Fig.~\ref{fig:mkp_modular_sketch}, to find the ideal kicker configuration for minimal magnet resources. On the other hand, parts of the septa from the PS Booster and parts from the present SPS injection system
can be reused to form a new movable septum conductor (also called \textit{blade}), whose second
half is 20 mm thick instead of today’s 40 mm for the whole blade. The reduced
thickness of the proposed MSI blades allows for more space for the beam envelope. Thus, this new MSI blade remains a fixed parameter.

\section{Hierarchical optimisation}

\begin{figure}[htb!]
\centering
\includegraphics[width=.95\columnwidth]{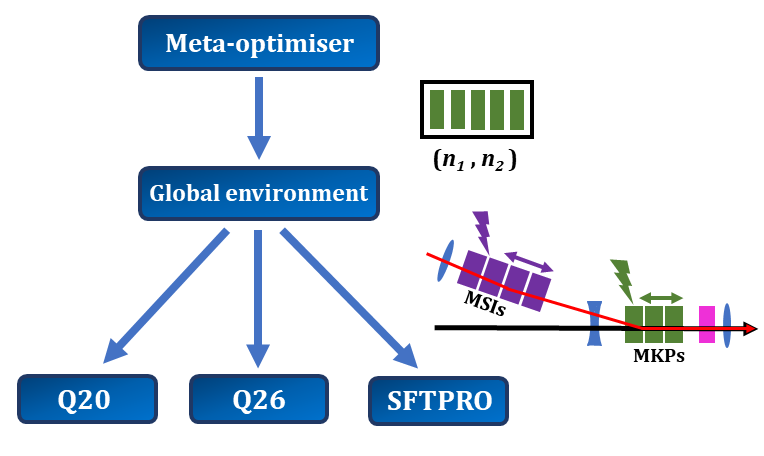}
\caption{\label{fig:optimiser_hierarchy} Illustration of the hierarchical optimiser.}
\end{figure}

The injection system requirements and actors spanning over multiple levels are incorporated in a hierarchical optimisation framework, where the low-level optimisation ensures a correct trajectory at injection, and the outer high-level optimisation minimises the magnet resources. An illustration of the flow across different hierarchy levels is shown in Fig.~\ref{fig:optimiser_hierarchy}. Numerical optimisation and machine learning are used in accelerator physics, and some attempts have also been made to exploit genetic algorithms (GA) for multi-objective optimisation to construct transfer lines~\cite{yann_genetic}, alas without a proposed final solution. 

The meta-optimiser (a wrapper function) first creates a global environment object with a given kicker magnet configuration from the tuple ($n_1$, $n_2$). The environment contains all the attributes and a \Verb|step| method, which takes the tuple as input parameter and outputs the high-level objective function to minimise, defined as 

\begin{equation}
\textrm{objective function} = \frac{1}{a_1} \bigg(\int Bdl \bigg)^2 + \frac{1}{a_2} B^2,
\label{eq:obj_func_high}
\end{equation}

where $a_i$ are normalisation constants. The global environment creates low-level environments for all the three optics (Q20, Q26 and SFTPRO). Each of these low-level environments contain the respective simulation process in MADX, and also a \Verb|step| method that takes magnet strengths and positions of the MSI and MKP, outputting the low-level objective function 

\begin{align}
\begin{split}
&\textrm{objective function} = \frac{1}{a_3} x^2 + \frac{1}{a_4} x'^2, \\&\textrm{penalty} = \textrm{ max}(0, A_{x\textrm{, min}} - A_{x})^2
\label{eq:objective_func_low}
\end{split}
\end{align}

to which a penalty is added: the square of the excess of the present acceptance $A_x$, with $A_{x\textrm{, min}}$ defined in Eq.~\eqref{eq:min_acc}. Thus, beam losses or an improperly dumped beam trajectory are heavily penalised. The low-level objective function with its penalties remains general, which can be replaced by or coupled with the vertical plane if desired. 

\section{Optimisation methods} 
\label{sec:optimisation_methods}

The low-level objective function is optimised for each optics with the adaptive Nelder-Mead algorithm, implemented in \Verb|scipy|~\cite{scipy}. The global environment then selects the magnet positions from the optics that requires the highest $\int~Bdl$, fixes these magnet positions and re-optimises their strengths - this approach guarantees that we first satisfy the case requiring the highest integrated field. In our study, the magnet strengths for the other optics could always be re-optimised with these new magnet positions.   

At the higher level, the optimiser evaluates the high-level objective function of the global environment and finds the ideal pair ($n_1$, $n_2$), using a GA with the library \Verb|pymoo|~\cite{pymoo}. GAs are inspired by the process of natural selection, where a starting population of initial settings is sampled and the (high-level) objective functions are evaluated. The fittest candidates survive and are selected for the crossover, combining into offspring where some mutations may occur according to pre-defined probabilities. Even with a starting population size of 10 randomly sampled pairs of settings, the algorithm converges already after 10-15 iterations to similar candidate solutions.

\section{Results} 
\label{sec:Results}

\begin{figure}[htb!]
\centering
\includegraphics[width=\columnwidth]{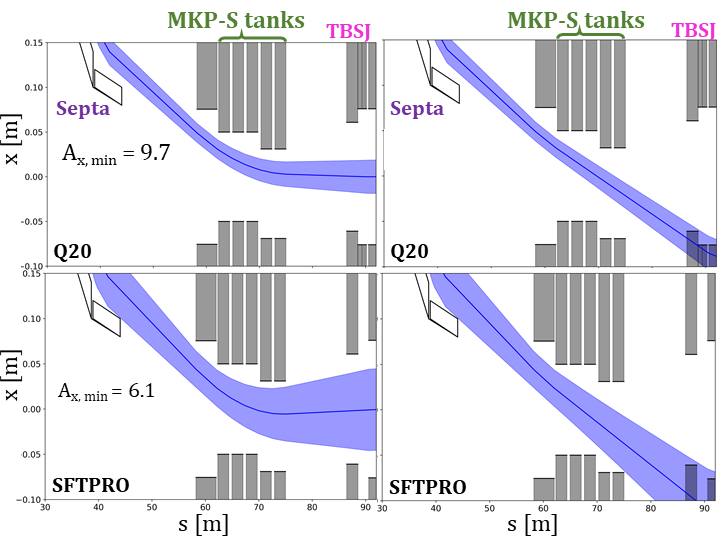}
\caption{\label{fig:hiearchical_simulation_results_compact} Optimisation results for Q20 and SFTPRO optics (left - injected beam, right - dumped beam).}
\end{figure}

The results produced by the hierarchical optimisation framework are presented in Fig.~\ref{fig:hiearchical_simulation_results_compact}, using the new thinner septum blades. The combination ($n_1 = 3$, $n_2 = 5$) yields the lowest value of the high- and low-level objective function values, and higher $A_{\textrm{x. min}}$ than today.

Although $\int B dl$ for Q20 and Q26 would increase by 5-6\%, the injected un-kicked beam reaches the TBSJ without the need of an extra dipole as used today - a great simplification for future operation. 

\begin{figure}[htb!]
\centering
\includegraphics[width=0.95\columnwidth]{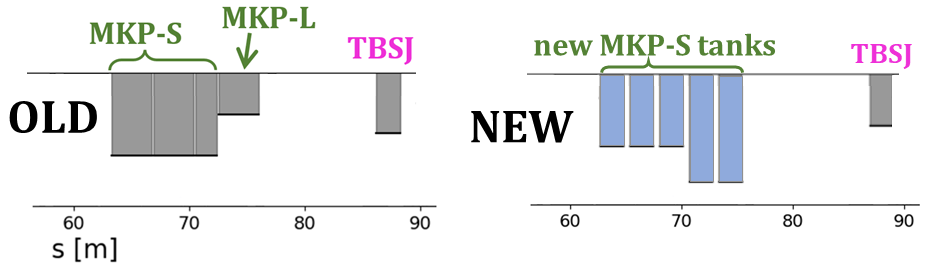}
\caption{\label{fig:new_and_old_mkp_configuration} Present and proposed MKP configuration.}
\end{figure}

A close-up of the MKP tanks in the present and proposed injection system is shown in Fig.~\ref{fig:new_and_old_mkp_configuration}. The new configuration only contains MKP-S type kickers, leading to less beam-induced heating and more uniform design and spare policy. The increased intermediate spacing for valves allows for independent vacuum sectors. The last two MKP tanks have been shifted by 1.9 cm horizontally in the new configuration to allow for more space to the dumped beam. The main drawback of this configuration is the fact that the external side of the last MKP tank is exposed to direct beam impact in case of self-triggering of the MKP while beam is circulating. More studies are needed to evaluate alternative solutions and to assess the risk associated with direct impact on the MKP, possibly with local shielding. 

\section{Conclusions} 
\label{sec:Conclusion}

The current SPS injection system has seen a radical evolution since its first design, and the characteristics of the injected beam have changed significantly. A full refactoring may be needed to solve long-standing issues, like the different kicker types and the bulky MSI. We presented a hierarchical optimisation framework that gradually constructs a new injection sequence and generates ideal candidate solutions for all SPS optics - finding the ideal balance between minimal magnet resources and dumped beam trajectory, while considering realistic hardware. The developed framework is rather generic and provides a basis for designing other accelerator systems, even with higher complexity. We stress that although the SPS injection is a specific case, the high-level discrete design of magnet modules and configurations can in principle be extended to almost any accelerator sequence, as described in a pseudo-algorithm in~\cite{elias_report}.


%
\bibliographystyle{plain}

%
%

\end{document}